\documentclass[prb,twocolumn,showpacs,aps,floatfix]{revtex4}
\usepackage{graphicx}

\begin{document}

\title{Influence of Band and Orbital Degeneracies
       on Ferromagnetism in the Periodic Anderson Model }

\author{Unjong Yu$^{1,2}$}
\author{Krzysztof Byczuk$^{1,3}$}
\author{Dieter Vollhardt$^1$}

\affiliation{$^1$Theoretical Physics III, Center for Electronic
  Correlations and Magnetism, Institute of Physics, University of
  Augsburg, D-86135 Augsburg, Germany\\
  $^2$Department of Physics and Astronomy, Louisiana State University,
  Baton Rouge, Louisiana 70803, USA \\
  $^3$Institute of Theoretical Physics,
  University of Warsaw, ul. Ho\.za 69, 00-681 Warszawa, Poland
}

\date{\today}

\begin{abstract}
We investigate the influence of degeneracies of the conduction band
and the $f$-orbital on the stability of ferromagnetism in the
periodic Anderson model. To this end we calculate the temperature
dependence of the inverse susceptibility for different degeneracies
$D^c, D^f$ and conduction electron densities $n^c$ within the
dynamical mean-field theory.
A strong increase of the Curie temperature $T_c$ with the degeneracy
$D^f$ of the localized $f$-level is found. For $D^c \leq D^f$ a
simple \emph{ansatz} based on a mean-field treatment of the RKKY
interaction is shown to imply a scaling behavior of $T_c$ as a
function of the conduction electron density per band which is well
obeyed by the numerical results. In particular, $T_c$ is found to
have a maximum at $n^c / D^c\approx 0.3$.
\end{abstract}

\pacs{71.10.Fd, 75.10.-b, 75.20.Hr}

\maketitle

\section{Introduction}

The periodic Anderson model (PAM) is the minimal model for the
investigation of interacting, localized electrons (e.g.,
$f$-electrons) hybridizing with a non-interacting band of conduction
electrons as in the rare earths and actinides. Indeed, the PAM can
explain many of the characteristic properties of heavy fermion
systems,\cite{Czycholl86} intermediate valence materials,\cite{Czycholl86,Varma76}
and Kondo insulators.\cite{Riseborough00} While it is well known that the PAM can
account for the long-range antiferromagnetic correlations at and
near half-filling of the conduction band, ferromagnetic phases of
this model have received less attention. Nevertheless, ferromagnetism
induced by
localized\cite{slaveB,Rasul00,Guerrero96,Tahvildar97,Santini99,Meyer00}
and even slightly delocalized\cite{Batista02} $f$-electrons is now
known to be a generic property of this model away from half-filling.
In fact, there exist a number of $f$-electron materials with
ferromagnetic phases, e.g., UCu$_2$Si$_2$,\cite{Chelmicki85}
UCu$_2$Ge$_2$,\cite{Chelmicki85} CeRh$_3$B$_2$,\cite{Malik85}
YbNiSn,\cite{Bonville92} URhSi,\cite{Tran98} URhGe,\cite{Tran98}
and CeRuPO.\cite{Krellner07}
Since the ions providing the $f$-electrons in these systems are
rather far apart, their hybridization with conduction electrons must
be considered essential for the stabilization of ferromagnetism.
This makes the PAM an appropriate model for the investigation of
these systems.

The conventional PAM does not take into account any degeneracies of
the electrons, e.g., the fact that without crystal field splitting
the $f$-orbitals have a seven-fold degeneracy. This degeneracy can
be expected to influence the magnetic properties of materials. For
example, in the case of the Hubbard model band-degeneracy is known
to greatly enhance ferromagnetism due to presence of Hund's rule
couplings.\cite{Vollhardt99,2Band2,2Band3} Therefore, a more
realistic study of ferromagnetism in the PAM should explicitly
consider band and orbital degeneracies of the electrons.\cite{Footnote1}
In this paper, we present a detailed study of the
influence of band degeneracy of the conduction band and/or  orbital
degeneracy of the localized levels on the stability of
ferromagnetism in the PAM for different conduction electron
densities. For this purpose we employ the dynamical mean-field
theory (DMFT) with quantum Monte-Carlo (QMC) as the impurity solver.
In particular, we show that, and explain why, $T_c$ increases with
the degeneracy of the $f$-level. The paper is organized as follows.
In Sec.~\ref{Model} we introduce the model and the calculation
method in detail. The result of the magnetization, magnetic
susceptibility, and phase diagrams in the PAM with degeneracy are
presented and discussed in Sec.~\ref{sec:result}. A summary is
presented in Sec.~\ref{Conclusion}.

\section{Model and method of calculation\label{Model}}

The multi-orbital PAM investigated here has the form
\begin{eqnarray}
H &=& - \sum_{\langle i,j \rangle  l \sigma} t^c_{l}
                   c^{\dagger}_{i l \sigma}  c^{}_{j l \sigma}
    + \sum_{i l \sigma} \varepsilon^{c}_{l} c^{\dagger}_{i l \sigma}  c^{}_{i l \sigma}
  \nonumber \\
    &&+ \sum_{i m \sigma} \varepsilon^{f}_{m} f^{\dagger}_{i m \sigma}  f^{}_{i m \sigma}
      + U \sum_{i m} n^{f}_{i m \uparrow} n^{f}_{i m \downarrow}
  \nonumber \\
    &&+ \sum_{i \sigma_1 \sigma_2} \sum_{m_1 < m_2} (U' - \delta_{\sigma_1 \sigma_2}F) ~
                     n^{f}_{i m_1 \sigma_1} n^{f}_{i m_2 \sigma_2}
  \nonumber \\
    &&+ V \sum_{i \sigma} \sum_{l m} \left( c^{\dagger}_{i l \sigma}
           f^{}_{i m \sigma} + {\rm H.c.} \right) ,
\end{eqnarray}
where $c^{\dagger}_{i l \sigma}$ ($f^{\dagger}_{i m \sigma}$)
creates conduction (localized) electrons with orbital $l$ ($m$) and spin
$\sigma$ at site $i$. Furthermore, $t^c_l$ is the
hopping parameter of the conduction electron, while
$\varepsilon^{c}_l$ and $\varepsilon^{f}_m$ denote the center of the
conduction band and the energy level of the $f$-electrons,
respectively. The orbital index
 can take the values $l=1, \dots, D^c$ for the conduction
band and $m=1, \dots, D^f$ for the $f$-level. In this paper, we
consider only degenerate bands, i.e., $t^c_l \equiv t^c$,
$\varepsilon^{c}_l \equiv \varepsilon^{c}$, and $\varepsilon^{f}_m
\equiv \varepsilon^{f}$. Furthermore, $U$ and $U'$ are the intra-
and inter-orbital Coulomb repulsion, respectively, $F$ is the Ising
component of the Hund's rule coupling, and
$V$ represents the hybridization between the conduction band and the
localized orbital. For the conduction band we assume a Bethe density
of states (DOS) with $\rho_c(\varepsilon) = (1/2\pi) \sqrt{4-\varepsilon^2}$
per spin. The bandwidth of the conduction band ($W=4$) defines the
energy scale. We note that the magnetic properties of the PAM do not
seem to depend on the specific shape of the non-interacting DOS,
since a Gaussian DOS gives qualitatively similar
results.\cite{Tahvildar97} In this paper we fix the interactions,
the hybridization, and the chemical potential at the values $U=1.5$,
$U'=1.1$, $F=0.2$, $V=0.6$, and $\mu = \varepsilon^f + U / 2 + (D^f
-1)(U'-F/2)$, respectively. For the latter choice of $\mu$, the
$f$-level is approximately half-filled,\cite{Footnote2} thus
permitting local moments to be formed.

In the local moment regime of the PAM ferromagnetism is due to the
magnetic interaction between the localized magnetic moments mediated
by the non-interacting conduction electrons. While a strong
hybridization increases the magnetic interaction, it simultaneously
weakens the magnetic moments. For this reason, static mean-field
theories or perturbative approaches do not describe ferromagnetic
solutions of the PAM adequately. Here the DMFT \cite{DMFT} has
proved to be a reliable non-perturbative investigation scheme. The
DMFT is based on the limit of infinite dimensions of the lattice,
where the many-body problem reduces to a local one and can, in
principle, be solved exactly.\cite{Metzner89}

The structure of the DMFT self-consistency equations for the PAM is
the same as that for the Hubbard model, except for the calculation
of the lattice Green function through the $\mathbf{k}$-integrated
Dyson equation,\cite{Jarrell93} which now reads
\begin{eqnarray}
& G^{c}_{l \sigma}(\omega_n) = \frac{1}{N_k} \sum\limits_{\mathbf{k}}
             \frac{1}{i\omega_n - \varepsilon^c_{l} - \varepsilon^{}_{l \mathbf{k}}
           - \sum\limits_{m} \frac{V^2}{i\omega_n - \varepsilon^f_{}
           - \Sigma_{m \sigma}(\omega_n)}} & ~~~~~~
           \\
& G^{f}_{m \sigma}(\omega_n) = \frac{1}{N_k} \sum\limits_{\mathbf{k}}
             \frac{1}{i\omega_n - \varepsilon^f_{} - \Sigma_{m \sigma}(\omega_n)
           - \sum\limits_{l} \frac{V^2}{i\omega_n - \varepsilon^c_l
           - \varepsilon^{}_{l \mathbf{k}}}}&\!\!.~~~~~~
\end{eqnarray}
Here $\Sigma_{m \sigma}(\omega_n)$ is the self-energy of the
$f$-electron with orbital $m$ and spin $\sigma$, and $N_k$ is the
number of $\mathbf{k}$ points in the summation. The Green functions
are represented as a function of the Matsubara frequency $\omega_n =
(2n+1) \pi /\beta$, where $\beta=1/k_B T$. Then, the bath Green
function is determined by the self-consistency condition ${\cal
G}_{m\sigma}^{-1}(\omega_n) = \Sigma_{m\sigma}(\omega_n) +
[G^f_{m\sigma}(\omega_n)]^{-1}$, and the Green function is obtained
by solving the effective single-impurity problem using
QMC.\cite{Hirsch86} This QMC-method involves a time discretization
$\Delta \tau = \beta / L$ and  subsequent extrapolation $\Delta \tau
\rightarrow 0$. In the present work we do not extrapolate in every
case, and mostly work with $\Delta \tau = 0.25$, because we are
mainly interested in the qualitative behavior of $T_c$. As will be
shown later the value of $T_c$ calculated with $\Delta \tau = 0.25$
is within a few percent of the value obtained by the extrapolation
$\Delta\tau \rightarrow 0$. For small $U$ (smaller than the bandwidth)
the resulting error is small. As $U$ increases, the error
increases correspondingly.

The magnetic susceptibility was calculated from the two-particle
correlation function.\cite{Ulmke95} It can also be obtained from
the magnetization in a weak magnetic field. We found that both
methods give the same results, although for the same computational
time the statistical error in the latter method is much larger.

\section{Results and discussion \label{sec:result}}

\begin{figure}[btp]
\includegraphics[width=7.8cm]{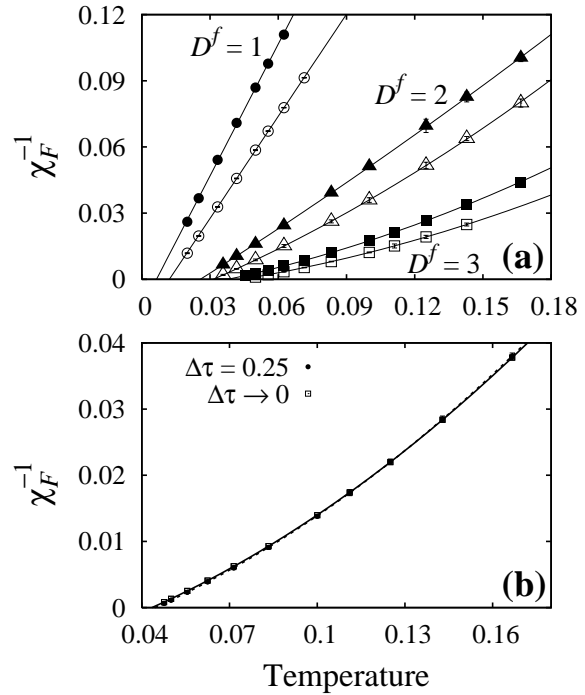}
\caption{\label{fig:sus_Mf}
  (a) Inverse susceptibility $\chi_F^{-1}$ of the PAM as a function of temperature
  for three values of the $f$-level degeneracy $D^f=1$ (circles), $D^f=2$ (triangles),
  $D^f=3$ (squares) and two values of the conduction electron
  density   $n^c=0.3$ (open symbols) and $n^c=0.5$ (filled symbols);
  the conduction band degeneracy is $D^c=1$ in all cases.
  Lines were obtained by fitting to Eq.~(\protect\ref{Eq:chi}).
  (b) Inverse susceptibility $\chi_F^{-1}$ calculated with  $\Delta\tau=0.25$
  compared with the extrapolated result ($\Delta\tau \rightarrow 0$) for
  $D^f=3$ and $n^c=0.4$.
}
\end{figure}

The numerical results obtained  for the inverse ferromagnetic
susceptibility $\chi_F^{-1}$ of the PAM for values of the $f$-level
degeneracy $D^f=1,2,3$ and electron densities $n^c=0.3,0.5$ are
shown in Fig.~\ref{fig:sus_Mf}(a). In the non-degenerate case ($D^f=1$),
the inverse susceptibility is found to be almost perfectly linear
for all temperatures. A linear behavior of the inverse
susceptibility near $T_c$, i.e., a Curie-Weiss mean-field behavior,
does not come unexpected in DMFT, although its validity was so far
proved only for the single-band Hubbard model.\cite{Byczuk02} The
numerical finding that this linear behavior persists up to high
temperatures, i.e., that the Curie-Weiss behavior merges into a
Curie law, is also not surprising in view of the low values of $T_c$
and a saturated magnetization in the magnetic phase.

For increasing degeneracy $D^f$, the inverse susceptibility becomes
more and more non-linear.
As can be seen from Fig.~\ref{fig:sus_Mf}(b), this is not due to the
finite value of $\Delta\tau$ since an extrapolation $\Delta\tau
\rightarrow 0$ does not change the result.
A non-linear behavior of the susceptibility is also seen in the
orbitally degenerate Hubbard model.\cite{Sakai06}
This non-linear behavior, i.e., the change of slope with
temperature, is due to the decrease of the effective magnetic moment
$\mu_{\rm eff}^{} = \sqrt{ \chi_F^{} \, (T-T_c) }$ with temperature.
This effect is negligible in the non-degenerate case ($D^f=1$). By
contrast, in the degenerate case the spins of the localized
electrons are aligned parallel due to the Hund's rule coupling near
$T_c$ (which is here found to be low) such that the effective
magnetic moment becomes large. On the other hand, at high
temperatures the effective magnetic moment decreases since the
electrons become independent and the spin of the localized electron
takes a random orientation on every site, owing to the thermal
fluctuation. Accordingly, the decrease of the effective magnetic
moment with increasing temperature leads to an increase of the slope
of the inverse susceptibility. This explains the upturn of the
inverse susceptibility for increasing temperature.

The non-linearity of $\chi_F^{-1}$ has to be taken into account in
the calculation of $T_c$ from the inverse susceptibility. Namely, a
purely linear fitting of $\chi_F^{-1}$ would lead to a serious error
in the value of $T_c$. We found the inverse susceptibility to be
well fitted by
\begin{eqnarray}
\chi_F^{-1} = A(T-T_c) + B(T-T_c)^2.
\label{Eq:chi}
\end{eqnarray}
The linear behavior is restored again at high temperatures, i.e.,
above the Fermi temperature (Curie law). The two linear regimes at
high temperatures and near $T_c$ are smoothly connected.

\begin{figure}[btp]
\includegraphics[width=8.5cm]{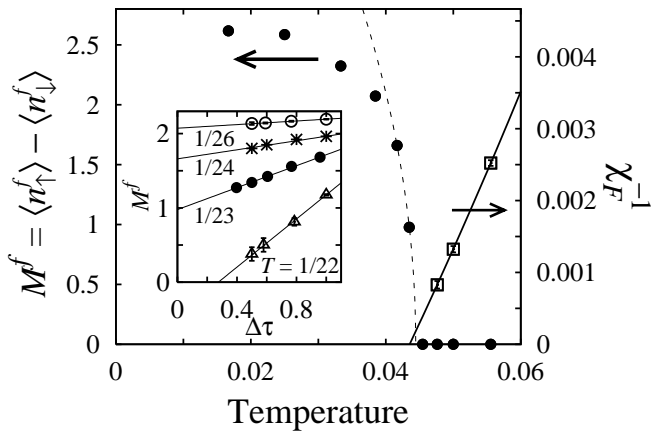}
\caption{\label{fig:M_Mf}
  Magnetization and inverse susceptibility $\chi_F^{-1}$
    extrapolated to $\Delta\tau \rightarrow 0$ as a function of temperature
    for $D^c=1$, $D^f=3$, and $n^c=0.4$. The dashed line marks
    a root-like disappearance of the magnetization fitted to the
    last two data points in the magnetic phase.
  The inset shows the $\Delta\tau$ dependence of the magnetization
  for  temperatures $T = 1/26, 1/24, 1/23$, and $1/22$.
}
\end{figure}

To check the validity of Eq.~(\ref{Eq:chi}) and the $T_c$ values
obtained from that expression, we calculated the magnetization
$M^f_{} = \langle n^f_{\uparrow} \rangle - \langle n^f_{\downarrow}
\rangle$ of the ferromagnetic phase of the degenerate PAM. To reduce
the error from thermal fluctuations in the QMC calculation we
performed an extrapolation $1/N_{\rm MC} \rightarrow 0$, where
$N_{\rm MC}$ is the number of Monte-Carlo steps in the QMC routine,
to calculate the magnetization at each iteration. Then we performed
an extrapolation $\Delta\tau \rightarrow 0$. As shown in the inset
of Fig.~\ref{fig:M_Mf}, the magnetization depends linearly on
$\Delta\tau$ and a spontaneous magnetization begins to appear at
temperature between $T=1/23$ and $1/22$. The magnetization and
inverse magnetic susceptibility obtained after the extrapolation
$\Delta\tau \rightarrow 0$ is presented in Fig.~\ref{fig:M_Mf}. The
value of $T_c$ obtained by fitting the susceptibility to
Eq.~(\ref{Eq:chi}) is clearly consistent with the values inferred
from the magnetization (inset of Fig.~\ref{fig:M_Mf}) and from the
root-fit to the magnetization (dashed curve in Fig.~\ref{fig:M_Mf}).

\begin{figure}[btp]
\includegraphics[width=7.8cm]{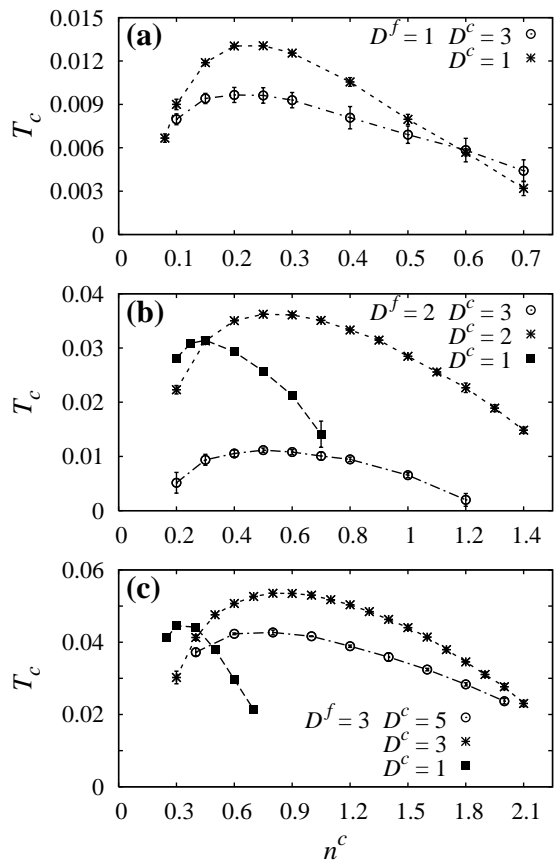}
\caption{\label{fig:Tc} Curie temperature as a function of
conduction electron density $n_c$ for different values of the
$f$-level degeneracy $D^f$ and conduction band degeneracy $D^c$. }
\end{figure}

The dependence of the Curie temperature $T_c$ on the conduction
electron density $n^c$ is shown in Fig.~\ref{fig:Tc} for several
values of the $f$-level ($D^f$) and conduction band ($D^c$)
degeneracy. The most remarkable feature is the pronounced increase
of $T_c$ with $D^f$. This behavior can be explained by the increase
of the local magnetic moment. Namely, the Hund's rule coupling
aligns the local magnetic moments on each site whereby the magnitude
of the local moment, $S$, increases with the degeneracy $D^f$; $T_c$
then increases accordingly. By contrast, the degeneracy of the
conduction band has a different effect: For increasing $D^c$ the
conduction electron density giving the maximal $T_c$
(i.e., the ``optimal" density) also
increases whereby the ferromagnetic phase expands to higher $n^c$
values. At the same time $T_c$ is hardly affected.
For $D^c \geq D^f$, however, $T_c$ almost always decreases
with increasing $D^c$.

\begin{figure}[btp]
\includegraphics[width=8.5cm]{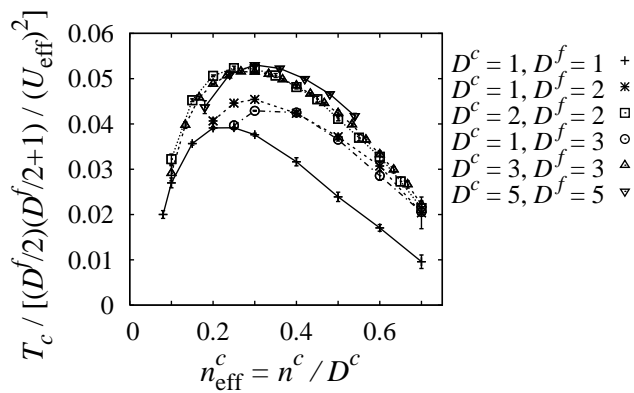}
\caption{\label{fig:Tc_rescale} Rescaled plot of the Curie
temperature $T_c$ \emph{vs.} conduction electron density $n_c$ for
several values of the $f$-level degeneracy $D^f$ and conduction band
degeneracy $D^c$, with $D^c < D^f$. }
\end{figure}

In order to explain the general dependence of $T_c$ on the
degeneracy shown in Fig.~\ref{fig:Tc} we modify the \textit{ansatz}
that we recently introduced in the case of the disordered PAM
without degeneracy,\cite{Yu08} which has the form $T_c(\mu) =
T_c^{0} F^f(\mu-\varepsilon^f) F^c(\mu-\varepsilon^c)$. Here the two
functions $F^f$ and $F^c$ describe the formation of the local
$f$-electron moments and the mediation of the magnetic ordering by
$c$-electrons, respectively. Since $\mu-\varepsilon^f$ is fixed,
$F^f(\mu-\varepsilon^f)$ depends only on the degeneracy. According
to the mean-field theory for the RKKY interaction in which this
\textit{ansatz} is indeed justified the local moment part ($F^f$) is
proportional to $J_{\rm eff}^2 \, S(S+1)$, where $J_{\rm eff}
=-8V^2/U_{\rm eff}$.\cite{multi_SW} The effective Coulomb
interaction $U_{\rm eff}$ is defined as $U_{\rm eff} = E(n_0^f+1) +
E(n_0^f-1) - 2 E(n_0^f)$, where $E(n_0^f)$ is the energy with the
$f$-electron density $n_0^f$.\cite{Anisimov91a,Anisimov91b} For a
half-filled $f$-level, $n_0^f=D^f$ and $U_{\rm eff} = U + (D^f-1)F$.
Thus, assuming $S=D^f/2$, $F^f$ is proportional to $(D^f/2)(D^f/2+1)
/ [U + (D^f-1)F]^2$. As in the conduction band part,
$F^c(\mu-\varepsilon^c)$ can be expressed by  $F^c(n^c_{\rm eff})$
with $n^c_{\rm eff}=n^c / D^c$, because $\mu-\varepsilon^c$ is
uniquely determined by the conduction electron density per band
($n^c_{\rm eff}$). Therefore, when we plot $T_c / \{(D^f/2)(D^f/2+1)
/ [U + (D^f-1)F]^2\}$ as a function of $n^c / D^c$, all curves
should fall onto the same curve. The result is shown in
Fig.~\ref{fig:Tc_rescale}. The deduced scaling behavior is seen to
be obeyed surprisingly well, given the simplicity of the assumptions
made in the derivation and the fact, that the parameter values
employed in our investigation are not limited to the RKKY regime.
Indeed $T_c$ increases rapidly with $n^c_{\rm eff}$ until it reaches
a maximum near $n^c_{\rm eff} = 0.3$ and then decreases slowly. The
maximal value of $T_c$ increases with $D^f$ and is about $0.05
\{(D^f/2)(D^f/2+1) / [U + (D^f-1)F]^2\}$.  In accordance with the
mean-field character of the above \emph{ansatz} the curves for
$D^c>1$ are found to fulfill the scaling behavior much better than
for $D^c=1$. For $D^c
> D^f$, which are not plotted in Fig.~\ref{fig:Tc_rescale}, the
above \emph{ansatz} does not apply. The additional conduction bands
seem to impede the magnetic ordering rather than mediate it.

\section{Conclusion \label{Conclusion}}

We computed the ferromagnetic susceptibility of the periodic
Anderson model (PAM), and from that the Curie temperature $T_c$, for
different values of the conduction band and $f$-level degeneracies
as well as electron densities $n^c$ within DMFT. The magnetic
susceptibility was found to deviate from the Curie-Weiss behavior
away from $T_c$, an effect which becomes stronger with increasing
degeneracy of the $f$-level. Our results show that the degeneracy
affects the ferromagnetic stability of the PAM strongly. In
particular, for $D^c \leq D^f$ the value of $T_c$ (i) increases with
the $f$-level degeneracy, (ii) approximately obeys a scaling law as
a function of the conduction electron density per band $n^c / D^c$,
which can be understood within a mean-field picture of the RKKY
interaction, and (iii) becomes maximal for $n^c / D^c\approx 0.3$.
This shows that realistic investigations of correlated electron
materials with localized $f$-levels must employ appropriate
generalizations of the PAM where the multi-orbital character of the
electrons is taken into account.

\begin{acknowledgments}
This work was supported in
part by the Sonderforschungsbereich 484 of the Deutsche
Forschungsgemeinschaft (DFG).
\end{acknowledgments}



\begin{thebibliography}{99}
\bibitem{Czycholl86} G.~Czycholl, Phys. Rep. \textbf{143}, 277 (1986);
          P.~Schlottmann, Phys. Rep. \textbf{181}, 1 (1989).
\bibitem{Varma76} C.~M.~Varma, Rev. Mod. Phys. \textbf{48}, 219 (1976).
\bibitem{Riseborough00} P.~S.~Riseborough, Adv. Phys. \textbf{49}, 257 (2000).
\bibitem{slaveB} A.~M.~Reynolds, D.~M.~Edwards, and A.~C.~Hewson,
         J.~Phys.:~Condens.~Matter \textbf{4}, 7589 (1992);
         B.~M\"{o}ller and P.~W\"{o}lfle,
         Phys. Rev. B \textbf{48}, 10320 (1993);
         R.~Doradzi\'{n}ski and J.~Spa\l{}ek, Phys. Rev. B
         \textbf{58}, 3293 (1998).

\bibitem{Guerrero96} M.~Guerrero and R.~M.~Noack, Phys. Rev. B
         \textbf{53}, 3707 (1996).

\bibitem{Tahvildar97} A.~N.~Tahvildar-Zadeh, M.~Jarrell, and J.~K.~Freericks,
         Phys. Rev. B \textbf{55}, R3332 (1997).
\bibitem{Santini99} P.~Santini, R.~L\'{i}manski, and P.~Erd\"{o}s, Adv. Phys. \textbf{48}, 537 (1999).
\bibitem{Rasul00} J.~W.~Rasul, Phys. Rev. B
         \textbf{61}, 15246 (2000).

\bibitem{Meyer00} D.~Meyer and W.~Nolting, Phys. Rev. B \textbf{62}, 5657 (2000).

\bibitem{Batista02} C.~D.~Batista, J.~Bon\v{c}a, and J.~E.~Gubernatis,
         Phys. Rev. Lett. \textbf{88}, 187203 (2002).
\bibitem{Chelmicki85} L.~Che\l{}micki, J.~Leciejewicz, and A.~Zygmunt,
         J. Phys. Chem. Solids \textbf{46}, 529 (1985).
\bibitem{Malik85} S.~K.~Malik, A.~M.~Umarji, G.~K.~Shenoy, P.~A.~Montano, and
         M.~E.~Reeves, Phys. Rev. B \textbf{31}, R4728 (1985).
\bibitem{Bonville92} P.~Bonville, P.~Bellot, J.~A.~Hodges, P.~Imbert,
         G.~J\'{e}hanno, G.~Le~Bras, J.~Hammann, L.~Leylekian,
         G.~Chevrier, P.~Thu\'{e}ry, L.~D'Onofrio, A.~Hamzic, and A.~Barth\'{e}l\'{e}my,
         Physica B \textbf{182}, 105 (1992).
\bibitem{Tran98} V.~H.~Tran, R.~Tro\'{c}, and G.~Andr\'{e},
         J. Magn. Magn. Mater. \textbf{186}, 81 (1998).
\bibitem{Krellner07} C.~Krellner, N.~S.~Kini, E.~M.~Br\"{u}ning, K.~Koch,
         H.~Rosner, M.~Nicklas, M.~Baenitz, and C.~Geibel,
         Phys. Rev. B \textbf{76}, 104418 (2007).

\bibitem{Vollhardt99} For a review see D.~Vollhardt, N.~Bl\"{u}mer,
K.~Held, M.~Kollar, J.~Schlipf, M.~Ulmke, and J.~Wahle,
Adv. in Solid State Phys. \textbf{38}, 383 (1999).

\bibitem{2Band2} A.~M.~Ole{\'{s}}, Phys. Rev. B \textbf{23}, 271 (1981);
G.~Stollhoff and P.~Thalmeier, Z. Phys. B \textbf{43}, 13 (1981);
G.~Stollhoff, A.~M.~Ole{\'{s}}, and V.~Heine, Phys. Rev. B \textbf{41}, 7028 (1990);
R.~Fr\'{e}sard and G.~Kotliar, Phys. Rev. B \textbf{56}, 12909 (1997).

\bibitem{2Band3} J.~Kuei and R.~T.~Scalettar, Phys. Rev. B \textbf{55}, 14968
(1997); J.~E.~Hirsch, Phys. Rev. B \textbf{56}, 11022 (1997); M.~Fleck,
A.~M.~Ole\'{s}, and L.~Hedin, Phys. Rev. B \textbf{56}, 3159 (1997).

\bibitem{Footnote1} The SU($N$) symmetric PAM,  where $N$ is the degeneracy of the
$f$-level and the conduction band, can be solved
exactly\cite{Coleman84} for $N \rightarrow \infty$, but does not
show a magnetic instability in this limit since the magnetic
interaction is suppressed\cite{Newns87} as $\sim 1/N^2$.

\bibitem{Coleman84} P.~Coleman, Phys. Rev. B \textbf{29}, 3035 (1984).
\bibitem{Newns87} D.~M.~Newns and N.~Read, Adv. Phys. \textbf{36}, 799 (1987).
\bibitem{Footnote2} Without hybridization this choice of
the chemical potential results in a half-filled $f$-level. For
non-zero hybridization, this does so only approximately.

\bibitem{DMFT} For reviews see A.~Georges, G.~Kotliar, W.~Krauth, M.~Rozenberg, Rev. Mod. Phys. {\bf 68}, 13
(1996); G.~Kotliar and D.~Vollhardt, Phys. Today \textbf{57}(3), 53
(2004).
\bibitem{Metzner89} W.~Metzner and D.~Vollhardt, Phys. Rev. Lett.
         \textbf{62}, 324 (1989).
\bibitem{Jarrell93} M.~Jarrell, H.~Akhlaghpour, and Th.~Pruschke,
         Phys. Rev. Lett. \textbf{70}, 1670 (1993).
\bibitem{Hirsch86} J.~E.~Hirsch and R.~M.~Fye, Phys. Rev. Lett.
         \textbf{56}, 2521 (1986).
\bibitem{Ulmke95} M.~Ulmke, V.~Jani\v{s}, and D.~Vollhardt, Phys. Rev. B
         \textbf{51}, 10411 (1995).
\bibitem{Byczuk02} K.~Byczuk and D.~Vollhardt, Phys. Rev. B
         \textbf{65}, 134433 (2002).
\bibitem{Sakai06} S.~Sakai, R.~Arita, K.~Held, and H.~Aoki,
         Phys. Rev. B \textbf{74}, 155102 (2006).
\bibitem{Yu08} U.~Yu, K.~Byczuk, and D.~Vollhardt, Phys. Rev. Lett.
        \textbf{100}, 246401 (2008).
\bibitem{multi_SW} This formula can be derived by a multi-orbital
        generalization of the Schrieffer-Wolff transformation; see
        B.~M\"{u}hlschlegel, Z. Phys. \textbf{208}, 94 (1968).
\bibitem{Anisimov91a} V.~I.~Anisimov and O.~Gunnarsson, Phys. Rev. B
        \textbf{43}, 7570 (1991).
\bibitem{Anisimov91b} V.~I.~Anisimov, J.~Zaanen, and O.~K.~Andersen, Phys. Rev. B
        \textbf{44}, 943 (1991).
\end{thebibliography}
\end{document}